\documentclass[aps,prl,twocolumn,superscriptaddress,floatfix]{revtex4-2}
\usepackage{amsmath,amssymb,bm,graphicx}
\usepackage[colorlinks=true,allcolors=blue]{hyperref}

\begin{document}

\title{Raman Circular Dichroism Reveals Higher-Order Quantum Geometry of Magnons}

\author{Yong Chan Kim}
\affiliation{Department of Applied Physics, Kyung Hee University, Yongin 17104, Republic of Korea}

\author{Youngsu Choi}
\affiliation{Department of Physics, Sungkyunkwan University, Suwon 16419, Republic of Korea}

\author{Kwang-Yong Choi}
\affiliation{Department of Physics, Sungkyunkwan University, Suwon 16419, Republic of Korea}

\author{Kyusung Hwang}
\email{kyusung.hwang@khu.ac.kr}
\affiliation{Department of Applied Physics, Kyung Hee University, Yongin 17104, Republic of Korea}

\date{\today}

\begin{abstract}
We develop a gauge-invariant framework that relates two-magnon Raman circular dichroism (RCD) to higher-order magnon quantum geometry. 
In the magnon band basis, the Raman operator decomposes into interband Berry connections, their covariant derivatives, and products of successive connections, generating higher-order, multi-state geometric tensors beyond the conventional single-band quantum metric and Berry curvature. Applying this framework to a field-polarized Kitaev magnet, we show that higher-order geometric tensors govern the dichroic response. Our results establish RCD as a spectroscopic probe of generalized magnon quantum geometry.
\end{abstract}

\maketitle

{\it Introduction} ---
Quantum geometry has become a central language for understanding topological and geometric responses of quantum materials.
While early studies focused on the Berry curvature~\cite{TKNN1982,Berry1984,Xiao2010,Nagaosa2010} and, more recently, the quantum metric~\cite{ProvostVallee1980,Resta2011,Torma2023,Gao2023Nature,Xu2023,Kang2025,Yang2025Science,Bernevig2022,Yan2025,Yang2025,Nagaosa2025,Kim2026,Queiroz2026}, current developments increasingly point to broader geometric structures involving momentum derivatives and multi-state processes, together with their manifestations in nonlinear transport and optical responses~\cite{SodemannFu2015,Yan2020,Ahn2020,Ahn2022,Agarwal2022,Slager2023,Ghorashi2024,Yan2024,Moore2025,Mitscherling2025,Froese2025,Schnyder2026}.
Extending this perspective to bosonic excitations, particularly magnons, is especially compelling. Magnons can possess nontrivial band geometry, but unlike electronic quasiparticles, their wave functions obey paraunitary Bogoliubov transformations~\cite{Katsura2010,Onose2010,Shindou2013,Murakami2014,Chisnell2015,Hirschberger2015,Mook2016,Chen2016,Hwang2020,Owerre2016,SKKim2016,Dai2018,Lee2020,Zhang2021,McClarty2022,Zhuo2025}. Recent studies have formulated single-band quantum geometry in this bosonic setting~\cite{Tesfaye2025}, uncovered its role in nonlinear thermal transport~\cite{Agarwal2023,Syljuaasen2025,Nasu2026}, and proposed spectroscopic probes of this geometry~\cite{Rubio2023,Kusminskiy2026,Slager2025,Zhang2026}. 
However, the higher-order, multi-state quantum geometry of magnons---and experimentally accessible probes thereof---remains largely unexplored.

In this paper, we show that Raman circular dichroism (RCD) provides a natural route to this problem.
Two-magnon RCD not only probes magnon topology~\cite{Rubio2023,Kusminskiy2026} but also reveals a hierarchy of higher-order geometric structures.
The key observation is that the Raman operator is decomposed into interband Berry connections (\(A\)), covariant derivatives (\(\nabla A\)), and products of Berry connections (\(AA\)).
Consequently, the dichroic Raman response separates into six geometric sectors
(\(AA\), \(AAA\), \(AAAA\), \(A\nabla A\), \(AA\nabla A\), and \(\nabla A\nabla A\)) revealing higher-order quantum-geometric tensors (QGTs).
Applying this framework to a field-polarized Kitaev magnet~\cite{Kitaev2006,Khaliullin2009,Rau2014,KimHS2015,Winter2016,Winter2017,Laurell2020,Maksimov2020,Li2022,Noh2026,Knolle2018,Takagi2019,Motome2020,Trebst2022,Matsuda2025,McClarty2018,Joshi2018,Chern2021,Ong2023,Chern2024,Loosdrecht2024,Perkins2026,Koller2026}, we demonstrate that the two-magnon RCD spectrum is shaped predominantly by higher-order geometric contributions and exhibits a characteristic sixfold field-angle dependence.
Our results establish RCD as a spectroscopic probe of generalized magnon quantum geometry beyond the conventional Berry curvature and quantum metric.

{\it Magnon Hamiltonian} ---
We consider generic quantum magnets described by the spin Hamiltonian
\begin{equation}
H =
\sum_{ij}
\mathbf S_i \!\cdot\! \mathbf J_{ij} \!\cdot\! \mathbf S_j
-
{\bf h} \cdot \sum_i \mathbf S_i ,
\end{equation}
where $\mathbf S_i$ is the spin operator at site $i$, the $3\times3$ matrix $\mathbf J_{ij}$ encodes the exchange interactions on bond $ij$, and ${\bf h}$ denotes an external magnetic field.
In a magnetically ordered state, magnon excitations can be analyzed using linear spin-wave theory (LSWT)~\cite{HP1940,Shindou2013,McClarty2022}, which yields the quadratic Hamiltonian
\begin{equation}
H_{\mathrm{LSW}} = {\rm const.} + \frac{1}{2}\sum_{\mathbf k}\Psi_{\mathbf k}^\dagger H_{\mathbf k}\Psi_{\mathbf k}.
\label{eq:H_LSW}
\end{equation}
Here, $\Psi_{\mathbf k}=[a_{1\mathbf k},\cdots,a_{N\mathbf k}|a^\dagger_{1,-\mathbf k},\cdots,a^\dagger_{N,-\mathbf k}]^T$ is the magnonic Nambu spinor of Holstein--Primakoff boson operators~\cite{HP1940}, and $H_{\mathbf k}$ is a $2N\times2N$ matrix describing magnon hopping and pairing; ${\bf k}$ denotes the crystal momentum and $N$ the number of magnetic sublattices.
The Bogoliubov transformation $\Psi_{\bf k}=T_{\bf k}\Gamma_{\bf k}$ maps the original magnon operators $\Psi_{\mathbf k}$ to the quasiparticle operators
$\Gamma_{\mathbf k}=
[
\gamma_{1\mathbf k},
\cdots,
\gamma_{N\mathbf k}|
\gamma^\dagger_{1,-\mathbf k},
\cdots,
\gamma^\dagger_{N,-\mathbf k}
]^T$
and diagonalizes the Hamiltonian matrix as
\begin{equation}
T_{\bf k}^\dagger H_{\mathbf k} T_{\bf k} = D_{\bf k}
={\rm diag} [\omega_{1\mathbf k},\cdots,\omega_{N\mathbf k}|\omega_{1,-\mathbf k},\cdots,\omega_{N,-\mathbf k}],
\end{equation}
where $\omega_{n{\bf k}}$ ($n=1,\cdots,N$) denotes the dispersion of the $n$-th magnon band.
To preserve the bosonic commutation relations, the $2N\times2N$ matrix $T_{\bf k}$ must satisfy the paraunitary condition~\cite{Shindou2013},
\begin{equation}
T_{\bf k}^\dagger I_{\rm B} T_{\bf k} = I_{\rm B}
~~~~{\rm or}~~~~T_{\bf k}^{-1} = I_{\rm B} T_{\bf k}^\dagger I_{\rm B},
\end{equation}
where $I_{\rm B}={\rm diag} [+1,\cdots,+1|-1,\cdots,-1]$ is the bosonic metric.

{\it Higher-order quantum geometric tensors} --- The fundamental building block of our theory is the interband Berry connection of the paraunitary magnon wave functions,
\begin{equation}
A^\mu_{mn}({\bf k})=\left[T_{\bf k}^{-1}i\partial_\mu T_{\bf k}\right]_{mn},
\label{eq:berry}
\end{equation}
where $\partial_\mu=\partial/\partial k_\mu$ and $m \ne n$.
Under the band-dependent gauge transformation $T_{\bf k} \rightarrow T_{\bf k}{\rm diag}[e^{i\phi_{1}({\bf k})},\cdots,e^{i\phi_{2N}({\bf k})}]$,
the interband connection transforms as $A^\mu_{mn}\rightarrow e^{i[-\phi_m({\bf k})+\phi_n({\bf k})]}A^\mu_{mn}$.
The simplest gauge-invariant tensor derived from the interband connection is the two-state QGT
\begin{equation}
Q^{\mu\nu}_{mn}=A^\mu_{nm}A^\nu_{mn},
\label{eq:two-state-QGT}
\end{equation}
whose real and imaginary parts define the two-state quantum metric and Berry curvature, respectively:
$g^{\mu\nu}_{mn}={\rm Re}Q^{\mu\nu}_{mn}$ and $\Omega^{\mu\nu}_{mn}=-2{\rm Im}Q^{\mu\nu}_{mn}$.
Summing these two-state quantities over the intermediate band recovers the conventional single-band quantum metric and Berry curvature~\cite{SM}.

We may construct further gauge-invariant multi-state QGTs:
\begin{align}
Q^{\rho\mu\nu}_{rmn}&=A^\rho_{nr}A^\mu_{rm}A^\nu_{mn},
\\
Q^{\sigma\rho\mu\nu}_{srmn}&=A^\sigma_{ns}A^\rho_{sr}A^\mu_{rm}A^\nu_{mn},
\label{eq:closed}
\end{align}
where the band indices $s,r,m,n$ are all distinct.
The closed-loop structure in band space (e.g., $n-s-r-m-n$) ensures the gauge invariance of these multi-state geometric tensors.

Covariant derivatives of the interband connection generate additional classes of gauge-invariant tensors:
\begin{align}
C^{\mu(\rho;\nu)}_{mn}&=A^\mu_{nm}\nabla_\rho A^\nu_{mn},
\\
B^{\mu\nu(\rho;\sigma)}_{rmn}&=A^\mu_{nr}A^\nu_{rm}\nabla_\rho A^\sigma_{mn},
\\
D^{(\rho;\mu)(\sigma;\nu)}_{mn}&=(\nabla_\rho A^\mu_{nm})(\nabla_\sigma A^\nu_{mn}),
\label{eq:cbd}
\end{align}
where the covariant derivative $\nabla A$ is defined by
\begin{equation}
\nabla_\sigma A^\nu_{mn}=(\partial_\sigma-iA^\sigma_{mm}+iA^\sigma_{nn})A^\nu_{mn}.
\label{eq:cov}
\end{equation}
The covariant derivative cancels the derivatives of the gauge-dependent phases and therefore transforms covariantly with the interband connection:
$\nabla_\sigma A^\nu_{mn}\rightarrow e^{i[-\phi_m({\bf k})+\phi_n({\bf k})]} \nabla_\sigma A^\nu_{mn}$.
This transformation property guarantees the gauge invariance of the three tensors.
The quantum connection $C^{\mu(\rho;\nu)}_{mn}$ and the third-order geometric tensor $Q^{\rho\mu\nu}_{rmn}$ have previously been shown to contribute to nonlinear optical responses in electronic systems~\cite{Ahn2022,Agarwal2022,Moore2025,Mitscherling2025,Froese2025}.
Here, we show that all the geometric tensors in Eqs.~(\ref{eq:two-state-QGT})--(\ref{eq:cbd}) arise naturally in the dichroic Raman response of magnons.

The geometric tensors can be evaluated numerically using gauge-invariant projector representations.
For magnons, we define the gauge-invariant projector $P_n({\bf k})=|u_{n{\bf k}}\rangle\langle\tilde u_{n{\bf k}}|$, where $|u_{n{\bf k}}\rangle$ is the $n$th column of $T_{\bf k}=[|u_{1{\bf k}}\rangle,\ldots,|u_{2N{\bf k}}\rangle]$ and $\langle\tilde u_{n{\bf k}}|$ is the $n$th row of $T^{-1}_{\bf k}$~\cite{Shindou2013,SM}.
These projectors obey $P_mP_n=\delta_{mn}P_n$, $\sum_nP_n=1$, and $P_n^\dagger=I_{\rm B}P_nI_{\rm B}$.
The interband connection and its covariant derivative can then be written as~\cite{Ahn2022}
\begin{align}
\hat e^\mu_{mn}
& \equiv A^\mu_{mn}|u_m\rangle\langle\tilde u_n|
=iP_m(\partial_\mu P_n)P_n ,
\\
\hat\nabla_\alpha\hat e^\mu_{mn}
&\equiv(\nabla_\alpha A^\mu_{mn})|u_m\rangle\langle\tilde u_n|
\nonumber\\
&=
i
P_m \Big[
(\partial_\alpha P_m) (\partial_\mu P_n)
+
\partial_\alpha \partial_\mu P_n
\Big] P_n .
\end{align}
Accordingly, the gauge-invariant tensors can be expressed as traces of projectors and their derivatives~\cite{Mitscherling2025}; for example,
\begin{align}
Q^{\rho\mu\nu}_{rmn}
&={\rm Tr}[\hat e^\rho_{nr}\hat e^\mu_{rm}\hat e^\nu_{mn}]
\\
&=-i\mathrm{Tr}\Big[
P_n(\partial_\rho P_r)(\partial_\mu P_m)(\partial_\nu P_n)
\Big],
\nonumber\\
C^{\mu(\rho;\nu)}_{mn}
&={\rm Tr}[\hat e^\mu_{nm}\hat\nabla_\rho\hat e^\nu_{mn}]
\\
&=
-\mathrm{Tr}
\Big[
P_n(\partial_\mu P_m)
\Big\{
(\partial_\rho P_m)(\partial_\nu P_n)
+
\partial_\rho\partial_\nu P_n
\Big\} \Big].
\nonumber
\end{align}
The complete set of projector representations is summarized in Table~\ref{tab:QGT-RCD} in End Matter, with their derivations provided in Supplemental Material~\cite{SM}.

\begin{figure*}[t]
\centering
\includegraphics[width=\linewidth]{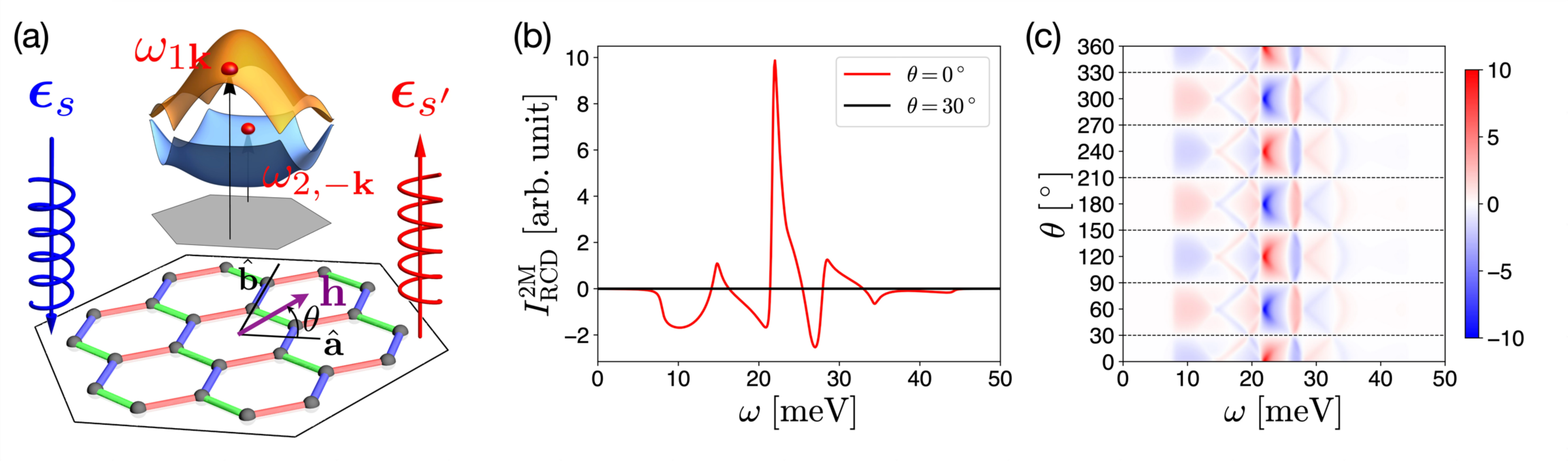}
\caption{Raman circular dichroism of the field-polarized Kitaev system.
(a) Schematic of Raman scattering.
The magnetic field ${\bf h}(\theta)$ is rotated within the honeycomb plane of the Kitaev system.
The yellow and blue surfaces depict the magnon energy bands over the first Brillouin zone.
(b) Two-magnon RCD intensity $I_{\rm RCD}^{\rm 2M}$ as a function of Raman energy shift $\omega$ obtained for the spin model~$H_{KJ\varGamma\varGamma' h}$ [Eq.~(\ref{eq:kgamma})] with the field aligned at $\theta=0^\circ$ (red) and $\theta=30^\circ$ (black).
(c) Field-angle dependence of $I_{\rm RCD}^{\rm 2M}(\omega)$.
The RCD intensity changes sign as the field direction crosses a bond direction ($\theta=30^\circ,90^\circ,150^\circ,210^\circ,270^\circ,330^\circ$) and vanishes when the field is aligned with a bond direction.
}
\label{fig:RCD-Kitaev}
\end{figure*}

\begin{figure*}[tb]
\centering
\includegraphics[width=\linewidth]{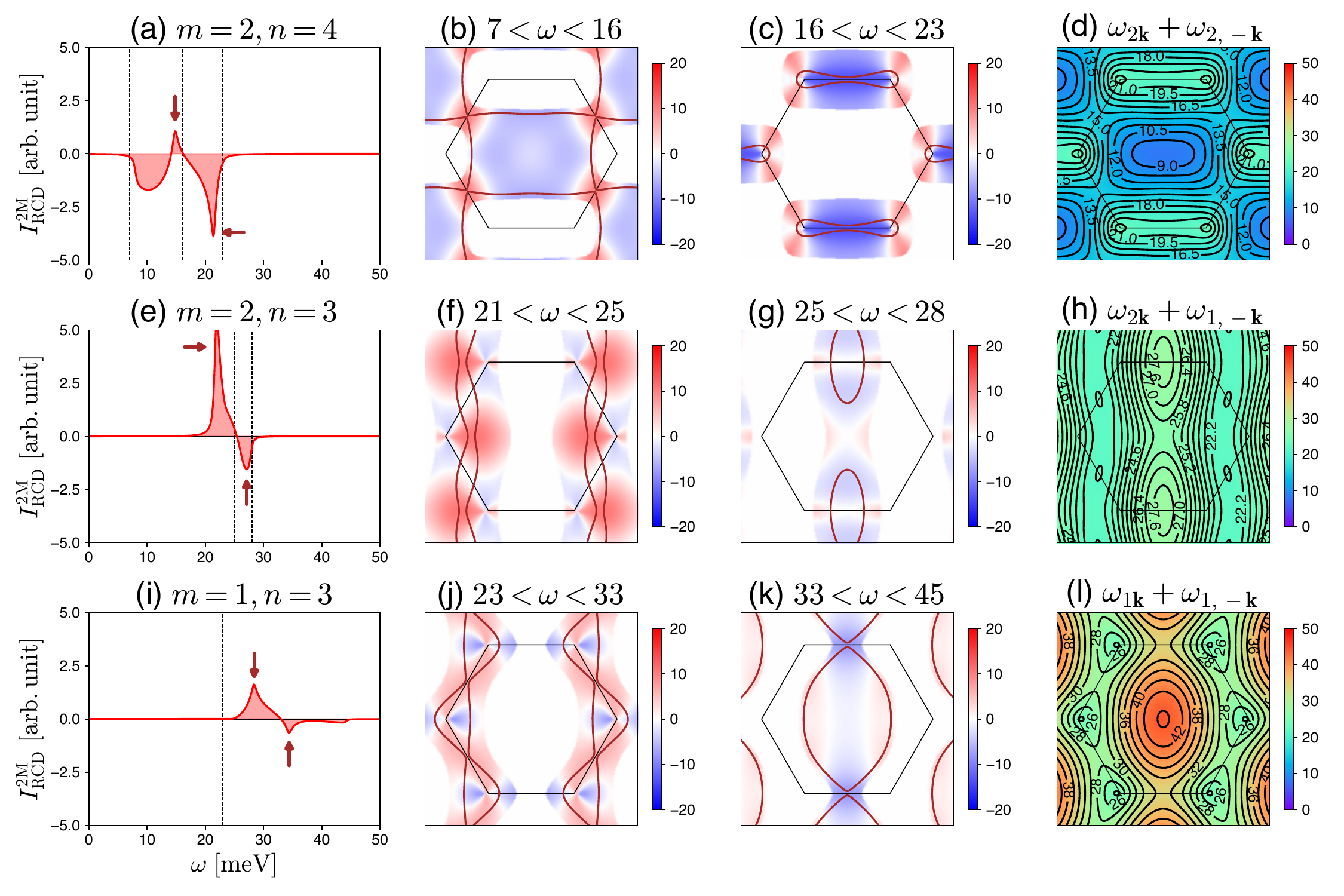}
\caption{Two-magnon RCD intensity $I_{\mathrm{RCD}}^{2\mathrm{M}}(\omega,\theta=0^\circ)$ resolved into three excitation sectors. The top, middle, and bottom rows correspond to $(m,n)=(2,4)$, $(2,3)$, and $(1,3)$, respectively. Panels (a), (e), and (i) show the sector-resolved spectra; (b), (c), (f), (g), (j), and (k) show momentum-space maps of $\mathcal{M}_{mn}(\mathbf{k})$ integrated over the indicated frequency windows; and (d), (h), and (l) show the corresponding two-magnon dispersions. The hexagons denote the first Brillouin zone, and the brown contours identify the momenta associated with the RCD peaks marked by arrows.}
\label{fig:RCD-decomposition}
\end{figure*}

\begin{figure}[tb]
\centering
\includegraphics[width=\linewidth]{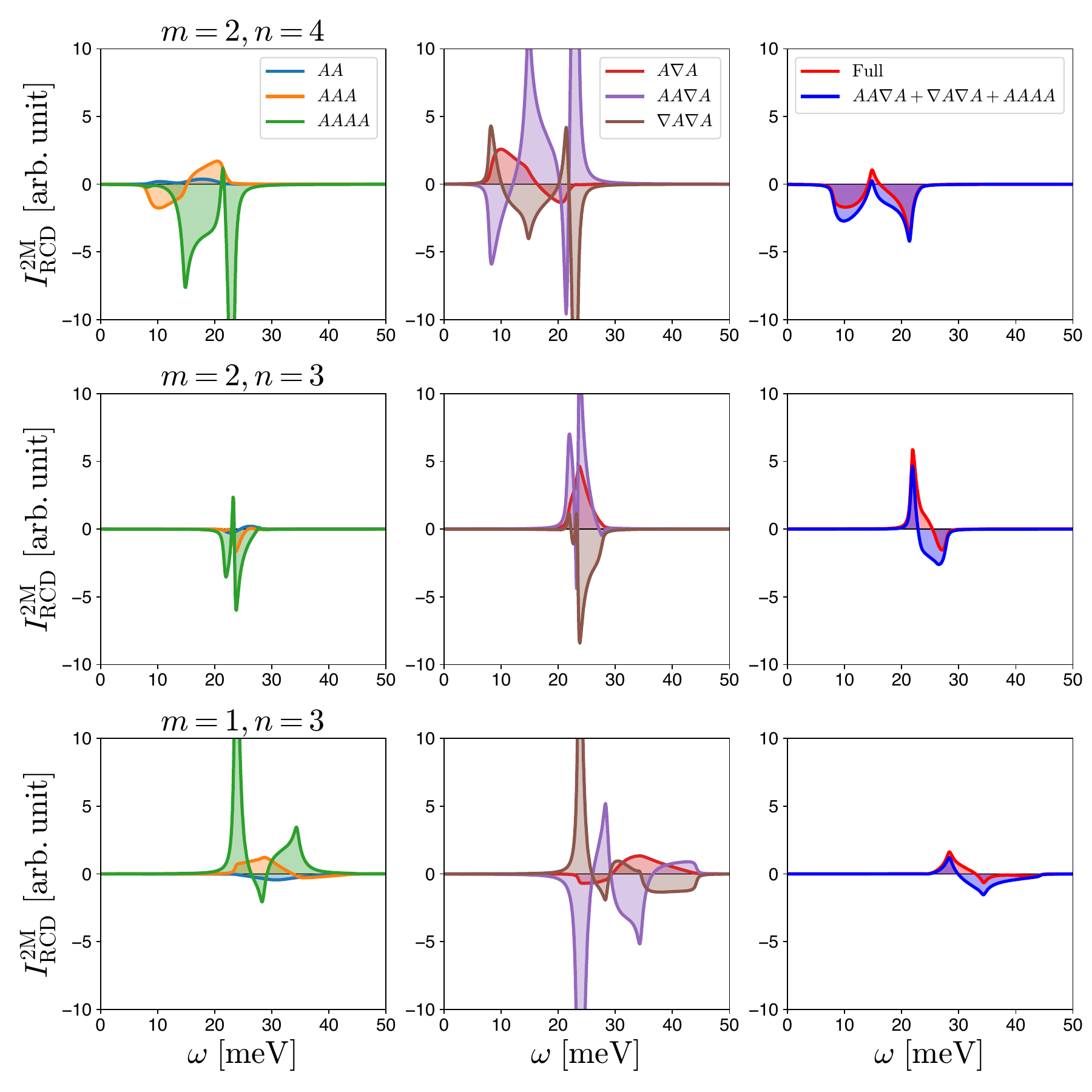}
\caption{Geometric decomposition of the two-magnon RCD intensity $I_{\mathrm{RCD}}^{2\mathrm{M}}(\omega,\theta=0^\circ)$ for $(m,n)=(2,4)$, $(2,3)$, and $(1,3)$ from top to bottom. The left and middle columns show the $AA$, $AAA$, $AAAA$ and $A\nabla A$, $AA\nabla A$, $\nabla A\nabla A$ contributions, respectively. The right column compares the full spectrum (red) with the sum of the dominant higher-order contributions, $AA\nabla A+\nabla A\nabla A+AAAA$ (blue).}
\label{fig:RCD-QGT}
\end{figure}

{\it Raman circular dichroism} ---
We now show how magnon quantum geometry enters the Raman response. To this end, we use the Loudon-Fleury-type Raman operator~\cite{LoudonFleury1968,Shastry1990,Devereaux2007,Perkins2021,Rubio2023,Kusminskiy2026,Perkins2026,Koller2026}
\begin{equation}
R_{ss'}=\sum_{ij}({\bm\epsilon}_s\cdot{\bf r}_{ij})({\bm\epsilon}_{s'}^*\cdot{\bf r}_{ij})
{\bf S}_i\cdot {\bf J}_{ij}\cdot{\bf S}_j,
\label{eq:lf}
\end{equation}
where ${\bf r}_{ij}$ is the lattice vector connecting sites $i$ and $j$, and $\boldsymbol{\epsilon}_s$ and $\boldsymbol{\epsilon}_{s'}$ are the polarization vectors of the incident and scattered light, respectively;
$\boldsymbol{\epsilon}_{\rm L,R}=(1,\pm i)/\sqrt{2}$ for the left- and right-circular polarizations.
The corresponding Raman scattering intensity is given by
\begin{equation}
I_{ss'}(\omega)
=
\frac{2\pi}{V} \sum_l
\left|\langle \psi_l \vert R_{ss'} \vert \psi_0 \rangle\right|^2
\delta\!\left(\omega - E_l + E_0\right),
\end{equation}
where $\omega$ is the Raman energy shift, $E_l$ and $\vert \psi_l \rangle$ ($l=0,1,\cdots$) are the energy eigenvalues and eigenstates, respectively, with $l=0$ denoting the ground state ($V$: volume of the system).
Using the magnon modes obtained within LSWT, we calculate the RCD intensity
\begin{equation}
I_{\mathrm{RCD}}(\omega)=I_{\rm LR}(\omega)-I_{\rm RL}(\omega).
\end{equation}

The quantum-geometric structure relevant to the present RCD response emerges in the two-magnon sector, for which the Raman operator takes the form
\begin{equation}
R^{\rm 2M}_{ss'}
\approx
-\frac{1}{2}\sum_{\mu,\nu}\epsilon_s^\mu\epsilon_{s'}^{\nu *}
\sum_{\bf k}\Gamma_{\bf k}^\dagger L^{\mu\nu}_{\bf k}\Gamma_{\bf k},
\end{equation}
where the $L$ matrix contains second derivatives of the Hamiltonian matrix,
\begin{equation}
L_\mathbf{k}^{\mu\nu}
=
T_{\mathbf{k}}^\dagger
\bigl(\partial_{\mu}\partial_{\nu}H_{\mathbf{k}}\bigr)
T_{\mathbf{k}} .
\end{equation}
We find that the $L$ matrix decomposes into three types of terms: interband connections ($A$), covariant derivatives ($\nabla A$), and successive interband connections ($AA$).
Specifically, the $L$-matrix element is given by
\begin{align}
L^{\mu\nu}_{mn}
&=\Big( i v^\mu_{mn}A^\nu_{mn}
+ i d_{mn}\nabla_\mu A^\nu_{mn}
+\sum_{r\ne m,n}d_{mnr}A^\mu_{mr}A^\nu_{rn} \Big)
\nonumber\\
&+\Big( \mu\leftrightarrow\nu \Big),
\label{eq:Ldecomp}
\end{align}
where $v^\mu_{mn}=\partial_\mu(d_m - [I_B]_m [I_B]_n d_n)$, $d_{mn}=(d_m - [I_B]_m [I_B]_n d_n)/2$, and $d_{mnr}=(- d_m - [I_B]_m [I_B]_n d_n)/2 + [I_B]_m [I_B]_r d_r$ with $d_r=[D_{\bf k}]_r$~\cite{SM}.
This finding allows us to express the two-magnon RCD intensity in terms of quantum geometric tensors:
\begin{align}
I^{\rm 2M}_{\rm RCD}(\omega)
&=\frac{2\pi}{V} \sum_{\bf k}\sum_{m=1}^{N}\sum_{n=N+1}^{2N}
{\rm Im}\Big[ (L^{xx}_{nm}-L^{yy}_{nm}) L^{xy}_{mn} \Big]
\nonumber\\
&~~~~~~~~~~~~~~~~~~~~~~
\delta(\omega-\omega_{m{\bf k}}-\omega_{n-N,-{\bf k}})
\label{eq:I_RCD}
\\
&=I^{\rm 2M}_{\rm RCD}\{AA\}+I^{\rm 2M}_{\rm RCD}\{AAA\}
\nonumber\\
&+I^{\rm 2M}_{\rm RCD}\{AAAA\}+I^{\rm 2M}_{\rm RCD}\{A\nabla A\}
\nonumber\\
&+I^{\rm 2M}_{\rm RCD}\{AA\nabla A\}+I^{\rm 2M}_{\rm RCD}\{\nabla A\nabla A\},
\label{eq:decomp}
\end{align}
where the second equality decomposes the response into six distinct geometric sectors.
Complete expressions for the six contributions in Eq.~(\ref{eq:decomp}) are provided in Supplemental Material~\cite{SM}.

{\it Two-magnon RCD of a field-polarized Kitaev magnet} ---
To illustrate our theory, we perform explicit calculations for a Kitaev spin system hosting topological magnons.
Specifically, we consider the $K$-$J$-$\varGamma$-$\varGamma'$ model in an in-plane magnetic field
\begin{align}
H_{KJ\varGamma\varGamma' h}
&=
\sum_{\langle jk \rangle_\gamma}
\Big[
K S_j^{\gamma} S_k^{\gamma} 
+ 
J \mathbf{S}_j \cdot \mathbf{S}_k 
+ 
\varGamma \big( S_j^{\alpha} S_k^{\beta} + S_j^{\beta}S_k^{\alpha}\big)
\nonumber\\
&+ 
\varGamma^\prime \big( S_j^{\alpha} S_k^{\gamma} +  S_j^{\gamma} S_k^{\alpha} + S_j^{\beta} S_k^{\gamma}+ S_j^{\gamma} S_k^{\beta}  \big) 
\Big]
\nonumber\\
&-{\bf h}(\theta)\cdot\sum_i{\bf S}_i.~~~~
\label{eq:kgamma}
\end{align}
This model has been proposed as a realistic description of the Kitaev material $\alpha$-RuCl$_3$~\cite{Rau2014,KimHS2015,Winter2016,Winter2017,Laurell2020,Maksimov2020,Li2022,Noh2026}.
We parametrize the in-plane magnetic field by ${\bf h}(\theta)=h\cos\theta\,\hat{\bf a}+h\sin\theta\,\hat{\bf b}$ and use the parameters
$K=-15~{\rm meV}$, $J=-0.1|K|$, $\varGamma=0.4|K|$, $\varGamma'=-0.02|K|$, and $h=g\mu_{\rm B} B=2\mu_{\rm B}\times14~{\rm T}$.
For these parameters, the spin moments align with the magnetic field; $\langle {\bf S}_i \rangle \propto {\bf h}(\theta)$.
Linear spin-wave theory yields two magnon bands ($N=2$) that are nearly degenerate near the corners of the Brillouin zone and carry Chern numbers $\pm 1$; see Fig.~\ref{fig:RCD-Kitaev}(a).

Figure~\ref{fig:RCD-Kitaev}(b) displays the calculated RCD intensity $I^{\rm 2M}_{\rm RCD}(\omega)$, which exhibits a broad continuum and several peaks at $\theta=0^\circ$ whereas it vanishes across all frequencies at $\theta=30^\circ$.
Figure~\ref{fig:RCD-Kitaev}(c) shows a color plot of the field-angle dependence of the RCD signals, revealing a distinct sixfold angular pattern.
To identify the origins of the RCD signals, we classify the two-magnon excitations $\omega_{m{\bf k}}+\omega_{n-N,-{\bf k}}$ into three sectors.
\begin{itemize}
\item $(m=2,n=4)$:
two magnons in the lower-energy band ($\omega_{2{\bf k}}+\omega_{2,-{\bf k}}$).
\item $(m=2,n=3)$ and $(m=1,n=4)$:
one magnon in the lower band and the other in the upper band ($\omega_{2{\bf k}}+\omega_{1,-{\bf k}}$)~\cite{Note1}.
\item $(m=1,n=3)$:
two magnons in the upper band ($\omega_{1{\bf k}}+\omega_{1,-{\bf k}}$).
\end{itemize}
Figures~\ref{fig:RCD-decomposition}(a), \ref{fig:RCD-decomposition}(e), and \ref{fig:RCD-decomposition}(i) show the RCD intensity resolved into these three sectors, respectively.
The corresponding two-magnon energy dispersions are plotted in Figs.~\ref{fig:RCD-decomposition}(d), \ref{fig:RCD-decomposition}(h), and \ref{fig:RCD-decomposition}(l).

Our calculations show that the RCD response is dominated by the $AA \nabla A$, $\nabla A \nabla A$, and $AAAA$ geometric sectors. Figure~\ref{fig:RCD-QGT} decomposes the RCD spectra into the six geometric components. The $AA$, $AAA$, and $A \nabla A$ contributions are comparatively weak, whereas the $AA \nabla A$, $\nabla A \nabla A$, and $AAAA$ terms 
make large contributions that reproduce the overall spectral profiles.
This comparison demonstrates that higher-order geometric contributions govern the dichroic response.

\begin{figure*}
\centering
\includegraphics[width=\linewidth]{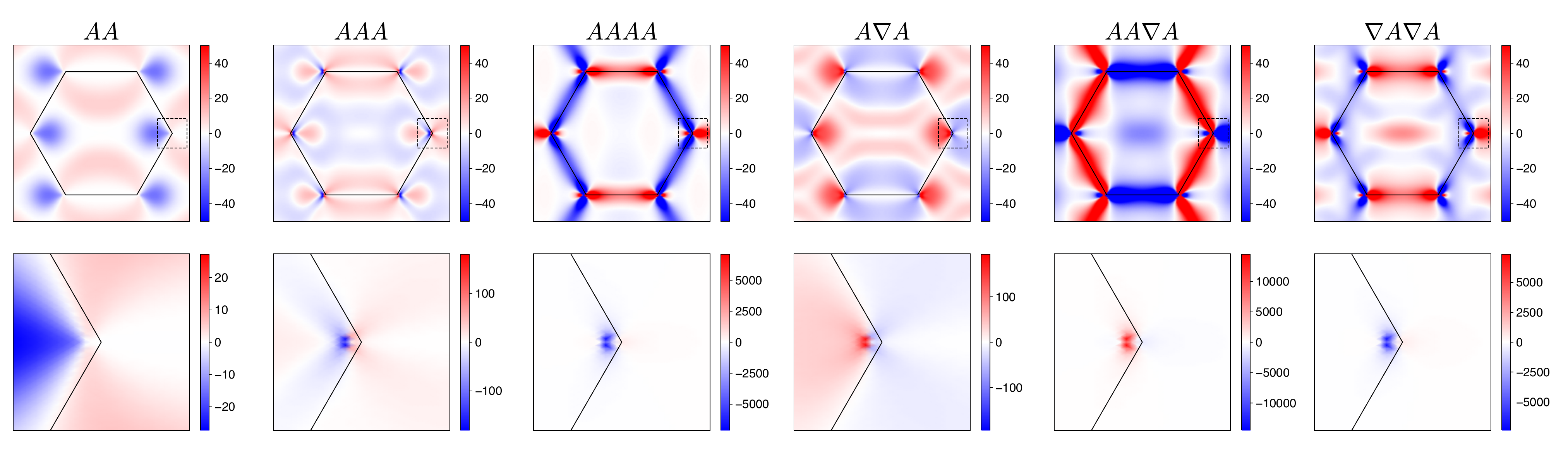}
\caption{The $L$-matrix element ${\rm Im}\Big[ (L^{xx}_{nm}-L^{yy}_{nm}) L^{xy}_{mn} \Big]$ for $m=2$ and $n=4$, decomposed into six geometric components.
Top: color plots of the six components restricted to the range $[-50,50]$.
Bottom: full-range plots of the square regions indicated in the top panels.}
\label{fig:matel24-QGT}
\end{figure*}

To precisely understand the geometric effects in the dichroic response,
we resolve the RCD intensity in momentum space by integrating the $L$-matrix elements over a finite energy window:
\begin{align}
&\mathcal{M}_{mn}({\bf k};E_1<\omega<E_2)
\\
&=
\int_{E_1}^{E_2} d\omega~~
{\rm Im}\Big[ (L^{xx}_{nm}-L^{yy}_{nm}) L^{xy}_{mn} \Big]
\delta(\omega-\omega_{m{\bf k}}-\omega_{n-N,-{\bf k}}) .
\nonumber
\end{align}
As shown in Fig.~\ref{fig:RCD-decomposition}(a), the $(m=2,n=4)$ sector contains a broad continuum and two peaks. Figures~\ref{fig:RCD-decomposition}(b) and \ref{fig:RCD-decomposition}(c) resolve, respectively, the continuum together with the lower-energy peak and the higher-energy peak.  Comparing these momentum-space maps with the $L$-matrix maps in Fig.~\ref{fig:matel24-QGT} of End Matter, we verify that the features in Fig.~\ref{fig:RCD-decomposition}(a) arise primarily from the $AA \nabla A$, $\nabla A \nabla A$, and $AAAA$ contributions, consistent with the spectral decomposition in Fig.~\ref{fig:RCD-QGT}.
Interestingly, the $L$-matrix maps exhibit sharp peaks associated with the near degeneracy of the magnon bands [Fig.~\ref{fig:matel24-QGT} (bottom)].
These peaks, however, cancel among the different geometric contributions, leaving no pronounced signature of the near degeneracy in Figs.~\ref{fig:RCD-decomposition}(b) and \ref{fig:RCD-decomposition}(c).
Thus, near-degenerate bands can strongly enhance individual geometric contributions without producing a corresponding feature in the observable RCD spectrum.
A comparison of Figs.~\ref{fig:RCD-decomposition}(a)--(d) allows us to trace
the broad continuum to two-magnon excitations near the zone center and the low- and high-energy peaks to excitations around the momenta marked by the brown lines.

\begin{figure*}
\centering
\includegraphics[width=\linewidth]{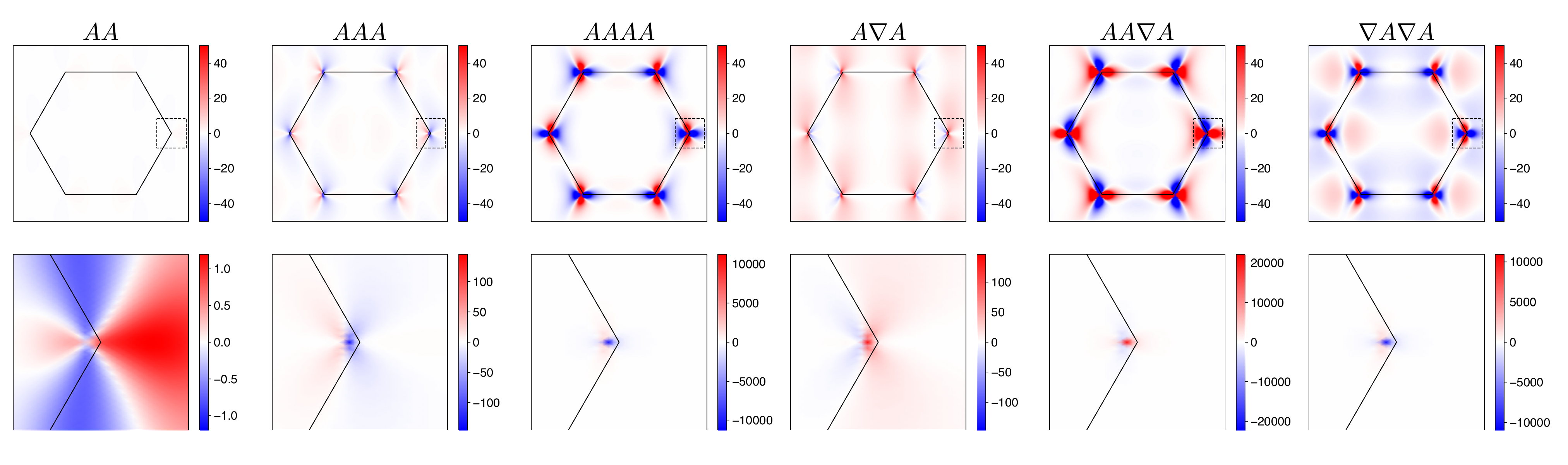}
\caption{The $L$-matrix element ${\rm Im}\Big[ (L^{xx}_{nm}-L^{yy}_{nm}) L^{xy}_{mn} \Big]$ for $m=2$ and $n=3$, decomposed into six geometric components.
Top: color plots of the six components restricted to the range $[-50,50]$.
Bottom: full-range plots of the square regions indicated in the top panels.}
\label{fig:matel23-QGT}
\end{figure*}

\begin{figure*}
\centering
\includegraphics[width=\linewidth]{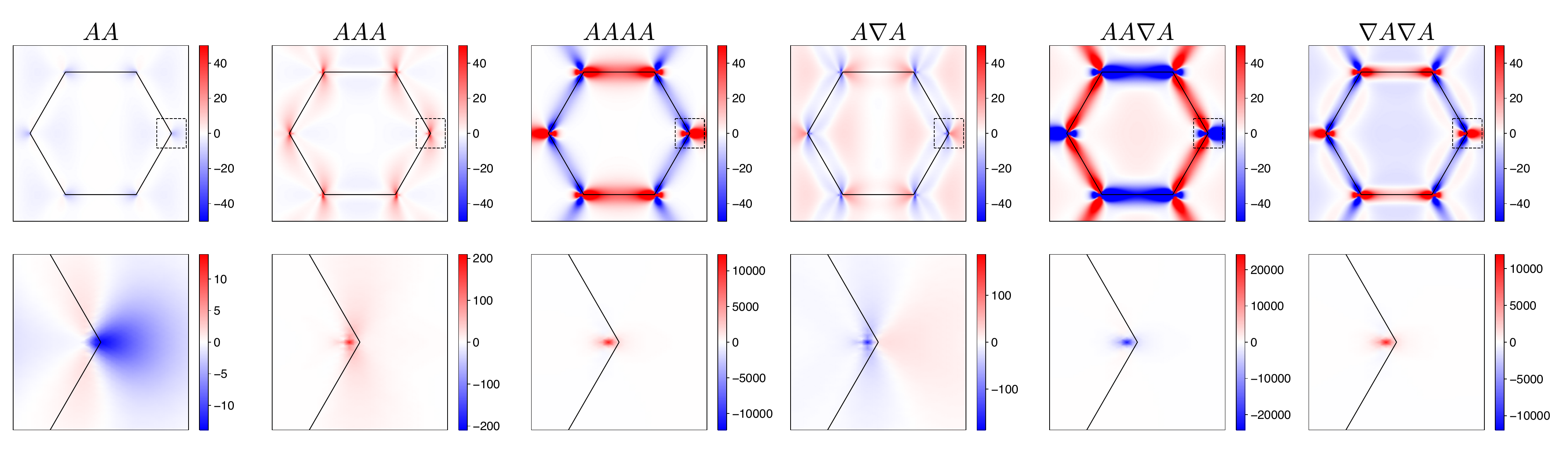}
\caption{The $L$-matrix element ${\rm Im}\Big[ (L^{xx}_{nm}-L^{yy}_{nm}) L^{xy}_{mn} \Big]$ for $m=1$ and $n=3$, decomposed into six geometric components.
Top: color plots of the six components restricted to the range $[-50,50]$.
Bottom: full-range plots of the square regions indicated in the top panels.}
\label{fig:matel13-QGT}
\end{figure*}

Applying the same analysis to the remaining sectors, we find that the peaks in Figs.~\ref{fig:RCD-decomposition}(e) and \ref{fig:RCD-decomposition}(i) are likewise dominated by the $AA \nabla A$, $\nabla A \nabla A$, and $AAAA$ contributions; see Figs.~\ref{fig:matel23-QGT} and \ref{fig:matel13-QGT} in End Matter. Comparing Figs.~\ref{fig:RCD-decomposition}(e)--\ref{fig:RCD-decomposition}(h) and Figs.~\ref{fig:RCD-decomposition}(i)--\ref{fig:RCD-decomposition}(l) allows us to identify the regions of the Brillouin zone responsible for these peaks. These regions lie away from the zone corners, where the two magnon bands become nearly degenerate; see the brown lines.

Lastly, we discuss the field-angle dependence of the RCD response [Fig.~\ref{fig:RCD-Kitaev}(c)].
The Kitaev system obeys the characteristic angular relations
\begin{align}
I_{\rm RCD}(\theta+120^\circ) &=I_{\rm RCD}(\theta),
\\
I_{\rm RCD}(180^\circ-\theta) &=-I_{\rm RCD}(\theta),
\end{align}
which follow from the $C_3$ and $C_2$ rotational symmetries of the zero-field Kitaev system~\cite{Hwang2022,Chern2024}.
The RCD intensity changes sign as the field direction crosses a bond direction ($\theta=30^\circ,90^\circ,150^\circ,210^\circ,270^\circ,330^\circ$) and vanishes when the field is aligned with a bond direction.
Because these relations are dictated by symmetry, they are not restricted to the two-magnon contribution but also apply to one-magnon and multi-magnon contributions.

{\it Discussion and outlook} ---
Raman circular dichroism has recently emerged as a probe of magnon-band topology~\cite{Rubio2023,Kusminskiy2026}, interaction-induced chiral phonons~\cite{Perkins2026}, and the quantum geometry of chiral spin liquids~\cite{Koller2026}. These studies have largely linked RCD to topology, chirality, and conventional geometric quantities such as the Berry curvature and quantum metric. Going beyond these developments, we show that the two-magnon RCD response encodes a hierarchy of higher-order, multi-state geometric tensors involving interband Berry connections and their covariant derivatives. In the field-polarized Kitaev magnet, these higher-order contributions dominate the RCD spectrum and produce a symmetry-enforced sixfold field-angle dependence. Our framework thus establishes RCD as a spectroscopic window into generalized magnon quantum geometry and opens a route to exploring higher-order geometric effects that remain largely uncharted in quantum magnetism.

{\it Acknowledgments} ---
We thank Kyung-Jin Lee, Myung Joon Han, Sung-Hyon Rhim, Heung-Sik Kim, Se Young Park, Beom Hyun Kim, and Hosub Jin for helpful discussions.
This work was supported by the National Research Foundation (NRF) of Korea under Grant No.~RS-2026-25475345 (Y.C.K. and K.H.), Grant No.~RS-2023-00214312 (Y.C.), and Grant No.~2020R1A5A1016518 (K.-Y.C.).

{\it Data availability} ---
The data supporting the findings of this study are available from the corresponding author upon reasonable request.


%

\section*{End Matter}

\begin{table*}
\begin{ruledtabular}
\begin{tabular}{lll}
Quantum geometric tensor & Projector representation & RCD component
\\
\hline
$Q_{mn}^{\mu\nu}
$
&
$
-
{\rm Tr}
\Big[
P_n (\partial_\mu P_m) (\partial_\nu P_n)
\Big]
$
&
$I_{\rm RCD}^{\rm 2M}\{AA\}$
\\
$Q_{rmn}^{\rho\mu\nu}$
&
$
-i
{\rm Tr}
\Big[
P_n
(\partial_\rho P_r)
(\partial_\mu P_m)
(\partial_\nu P_n)
\Big]
$
&
$I_{\rm RCD}^{\rm 2M}\{AAA\}$
\\
$Q_{srmn}^{\sigma\rho\mu\nu}$
&
$
\mathrm{Tr}
\Big[
P_n(\partial_\sigma P_s)(\partial_\rho P_r)(\partial_\mu P_m)(\partial_\nu P_n)
\Big]
$
&
$I_{\rm RCD}^{\rm 2M}\{AAAA\}$
\\
\hline
$C_{mn}^{\mu(\rho;\nu)}$
&
$
-
{\rm Tr}
\Big\{
P_n (\partial_\mu P_m) \Big[
(\partial_\rho P_m) (\partial_\nu P_n)
+
\partial_\rho \partial_\nu P_n
\Big]
\Big\}
$
&
$I_{\rm RCD}^{\rm 2M}\{A\nabla A\}$
\\
$B_{rmn}^{\mu\nu(\rho;\sigma)}$
&
$
-i
{\rm Tr}
\Big\{
P_n (\partial_\mu P_r)
(\partial_\nu P_m)
\Big[
(\partial_\rho P_m) (\partial_\sigma P_n)
+
\partial_\rho \partial_\sigma P_n
\Big]
\Big\}
$
&
$I_{\rm RCD}^{\rm 2M}\{A A\nabla A\}$
\\
$D_{mn}^{(\rho;\mu)(\sigma;\nu)}$
&
$
-
{\rm Tr}
\Big\{
P_n
\Big[
(\partial_\rho P_n) (\partial_\mu P_m)
+
\partial_\rho \partial_\mu P_m
\Big]
P_m
\Big[
(\partial_\sigma P_m) (\partial_\nu P_n)
+
\partial_\sigma \partial_\nu P_n
\Big]
\Big\}
$
&
$I_{\rm RCD}^{\rm 2M}\{\nabla A\nabla A\}$
\end{tabular}
\end{ruledtabular}
\caption{Quantum geometric tensors probed by magnon Raman circular dichroism.
The middle column shows the projector representations of the geometric tensors.}
\label{tab:QGT-RCD}
\end{table*}

\clearpage
\onecolumngrid

\setcounter{section}{0}
\setcounter{subsection}{0}
\setcounter{subsubsection}{0}
\setcounter{equation}{0}
\setcounter{figure}{0}
\setcounter{table}{0}
\renewcommand{\theequation}{S\arabic{equation}}
\renewcommand{\thefigure}{S\arabic{figure}}
\renewcommand{\thetable}{S\arabic{table}}

\begin{center}
{\large\bfseries Supplemental Material for\\
``Raman Circular Dichroism Reveals Higher-Order Quantum Geometry of Magnons''\par}
\vspace{1em}
Yong Chan Kim$^{1}$, Youngsu Choi$^{2}$, Kwang-Yong Choi$^{2}$, and Kyusung Hwang$^{1}$\\
\vspace{0.5em}
{\itshape
$^{1}$Department of Applied Physics, Kyung Hee University, Yongin 17104, Republic of Korea\\
$^{2}$Department of Physics, Sungkyunkwan University, Suwon 16419, Republic of Korea}
\end{center}

\vspace{1em}

\section{Single-state quantum geometry}

For magnon systems, the single-state quantum geometric tensor is defined as
\begin{align}
Q_{n{\bf k}}^{\mu\nu}
&=
\sum_{1 \le m(\ne n) \le 2N}
[A_{\bf k}^{\mu}]_{nm} [A_{\bf k}^\nu]_{mn}
\label{eq:single-band-QGT1}
\\
&=
\sum_{1 \le m(\ne n) \le 2N}
[T_{\bf k}^{-1} i\partial_{\mu} T_{\bf k}]_{nm} [T_{\bf k}^{-1} i\partial_{\nu} T_{\bf k}]_{mn}
\\
&=
\sum_{1 \le m(\ne n) \le 2N}
\langle \tilde{u}_{n{\bf k}} |
i \partial_\mu
|u_{m{\bf k}}\rangle
\langle \tilde{u}_{m{\bf k}} |
i \partial_\nu
|u_{n{\bf k}}\rangle ,
\end{align}
where $|u_{n{\bf k}}\rangle$ is the $n$th column of the Bogoliubov transformation matrix $T_{\bf k}$ and $\langle\tilde u_{n{\bf k}}|$ is the $n$th row of its inverse $T^{-1}_{\bf k}$ [see Eq.~(\ref{eq:T-matrices})].

The quantum geometric tensor can be recast in a more familiar form in terms of magnon velocities.
Differentiating the energy eigenvalue matrix yields an equation
\begin{equation}
\partial_{\mu}D_{\bf k}
=
\partial_\mu (T_{\bf k}^\dagger H_{\bf k} T_{\bf k})
~~~\Rightarrow~~~
V_{\bf k}^\mu
=
\partial_{\mu}D_{\bf k}
+
\left(
i D_{\bf k} A_{\bf k}^\mu + {\rm H.c.}
\right),
\label{eq:dH}
\end{equation}
where $V_{\bf k}^\mu$ is the velocity matrix,
\begin{align}
V_{\bf k}^\mu = T_{\bf k}^\dagger (\partial_\mu H_{\bf k}) T_{\bf k} .
\end{align}
The Berry connection matrix also satisfies
\begin{align}
A_{\bf k}^{\mu\dagger}=I_{\rm B}A_{\bf k}^{\mu}I_{\rm B}.
\label{eq:A_HC}
\end{align}
Combining Eqs.~(\ref{eq:dH}) and (\ref{eq:A_HC}) gives the interband connection in terms of the velocity matrix:
\begin{align}
[A_{\bf k}^\mu]_{mn}
=
\frac{-i[I_{\rm B}V_{\bf k}^\mu]_{mn}}{[I_{\rm B}D_{\bf k}]_m-[I_{\rm B}D_{\bf k}]_n}~~~(m \ne n).
\end{align}
Substituting this expression into Eq.~(\ref{eq:single-band-QGT1}) yields
\begin{align}
Q_{n{\bf k}}^{\mu\nu}
=
\sum_{m(\ne n)}
\frac{[I_{\rm B}V_{\bf k}^\mu]_{nm} [I_{\rm B}V_{\bf k}^\nu]_{mn}}{\left( [I_{\rm B}D_{\bf k}]_m - [I_{\rm B}D_{\bf k}]_n \right)^2} .
\label{eq:single-band-Q}
\end{align}
From this expression, we identify the quantum metric
\begin{align}
g_{n{\bf k}}^{\mu\nu}
= {\rm Re} [ Q_{n{\bf k}}^{\mu\nu} ]
=
\frac{1}{2}
\sum_{m(\ne n)}
\frac{[I_{\rm B}V_{\bf k}^\mu]_{nm} [I_{\rm B}V_{\bf k}^\nu]_{mn} + (\mu \leftrightarrow \nu)}{\left( [I_{\rm B}D_{\bf k}]_m - [I_{\rm B}D_{\bf k}]_n \right)^2} ,
\label{eq:single-band-g}
\end{align}
and the Berry curvature,
\begin{align}
\Omega_{n{\bf k}}^{\mu\nu}
= -2 {\rm Im} [ Q_{n{\bf k}}^{\mu\nu} ]
=
i
\sum_{m(\ne n)}
\frac{[I_{\rm B}V_{\bf k}^\mu]_{nm} [I_{\rm B}V_{\bf k}^\nu]_{mn} - (\mu \leftrightarrow \nu)}{\left( [I_{\rm B}D_{\bf k}]_m - [I_{\rm B}D_{\bf k}]_n \right)^2} .
\label{eq:single-band-Omega}
\end{align}
Together, these tensors form the quantum geometric tensor
\begin{align}
Q_{n{\bf k}}^{\mu\nu}
=
g_{n{\bf k}}^{\mu\nu}
-i\frac{1}{2}
\Omega_{n{\bf k}}^{\mu\nu},
\label{eq:single-band-QGT}
\end{align}
whose effects have recently been investigated in various bosonic systems~\cite{Rubio2023,Agarwal2023,Tesfaye2025,Syljuaasen2025,Slager2025,Zhang2026,Nasu2026}.

\section{Projector representations of geometric tensors}

Gauge-invariant geometric tensors are constructed from products of interband connections and their covariant derivatives.
Projector representations are particularly useful for numerical evaluation because they contain no gauge-dependent phase factors~\cite{Mitscherling2025}.

To construct these representations, we decompose the Bogoliubov transformation matrix and its inverse into column and row vectors, respectively:
\begin{align}
T_{\bf k}
=
\Big[ |u_{1{\bf k}}\rangle, \cdots, |u_{2N{\bf k}}\rangle \Big]
~~~~{\rm and}~~~~
T_{\bf k}^{-1}
=
I_{\rm B} T_{\bf k}^\dagger I_{\rm B}
=
\left[
\begin{array}{c}
\langle \tilde{u}_{1{\bf k}} |
\\
\vdots
\\
\langle \tilde{u}_{2N{\bf k}} |
\end{array}
\right] ,
\label{eq:T-matrices}
\end{align}
where $|u_{n{\bf k}}\rangle$ is a column of $T_{\bf k}$ and
$\langle \tilde{u}_{n{\bf k}} |$ is a row of $T_{\bf k}^{-1}$.
In this basis, the interband connection is written as
$
A_{mn}^\mu({\bf k})
=
\left[ T_{\bf k}^{-1} i \partial_\mu T_{\bf k} \right]_{mn}
=
\langle \tilde{u}_{m{\bf k}} |
i \partial_\mu
|u_{n{\bf k}}\rangle ,
$
and the projector onto the $n$th state is defined by
\begin{align}
P_n({\bf k}) = | {u}_{n{\bf k}} \rangle \langle \tilde{u}_{n{\bf k}} | ,
\end{align}
which is manifestly gauge invariant.
The projectors obey the following properties:
(i) $P_n^\dagger = I_{\rm B} P_n I_{\rm B}$,
(ii) $P_mP_n=0~(m\ne n)$,
and (iii) $P_nP_n=P_n$.
These relations yield several useful identities:
\begin{equation}
P_mP_n=0
~~~\Rightarrow~~~
P_m (\partial_\mu P_n) = - (\partial_\mu P_m) P_n,
\label{eq:id1}
\end{equation}
\begin{align}
P_nP_n=P_n
~~~&\Rightarrow~~~
P_n (\partial_\mu P_n) = (\partial_\mu P_n) (1-P_n)
\label{eq:id2}
\\
~~~&\Rightarrow~~~
P_n (\partial_\mu P_n) P_n = (\partial_\mu P_n) (1-P_n) P_n = 0,
\label{eq:id3}
\end{align}
\begin{equation}
P_m (\partial_\mu P_n) P_n
=
P_m (1-P_n)(\partial_\mu P_n)
=
P_m(\partial_\mu P_n).
\label{eq:id4}
\end{equation}
Here and below, we assume $m \ne n$.

To derive the projector representations, it is convenient to introduce the following matrix forms of the interband connection and its covariant derivative:
\begin{align}
\hat{e}_{mn}^\mu
&= A^\mu_{mn}|u_m\rangle\langle\tilde u_n|
\\
&= |u_m\rangle  \langle \tilde{u}_{m} |
i \partial_\mu
|u_{n}\rangle  \langle\tilde u_n|
\\
&= P_m (i \partial_\mu) P_n ,
\label{eq:A}
\end{align}
\begin{align}
\hat{\nabla}_\alpha \hat{e}_{mn}^\mu
&=
(\nabla_\alpha A^\mu_{mn})|u_m\rangle\langle\tilde u_n|
\\
&=
|u_m\rangle\langle\tilde u_m|
\Big[
\partial_\alpha (
A^\mu_{mn}|u_m\rangle\langle\tilde u_n|
)
\Big]
|u_n\rangle\langle\tilde u_n|
\\
&=
P_m (\partial_\alpha \hat{e}_{mn}^\mu) P_n
\\
&=
i
P_m \Big\{ \partial_\alpha [P_m (\partial_\mu P_n)] \Big\} P_n
\\
&=
i
P_m \Big[
(\partial_\alpha P_m) (\partial_\mu P_n)
+
(\partial_\alpha \partial_\mu P_n)
\Big] P_n .
\label{eq:Del-A}
\end{align}
This type of representation was introduced in Ref.~\cite{Ahn2022} to study optical responses in electronic systems.
We now use these expressions to construct projector representations of higher-order quantum geometric tensors.

\subsection{Second-order quantum geometric tensor}

In the projector representation, the second-order geometric tensor is
\begin{align}
Q_{mn}^{\mu\nu}
=
A_{nm}^{\mu} A_{mn}^\nu
=
{\rm Tr}
\Big[
\hat{e}_{nm}^{\mu} \hat{e}_{mn}^\nu
\Big]
&=
-
{\rm Tr}
\Big[
P_n (\partial_\mu P_m)
P_m (\partial_\nu P_n)
\Big]
\\
&=
-
{\rm Tr}
\Big[
P_n (\partial_\mu P_m) (\partial_\nu P_n)
\Big],
\end{align}
where Eqs.~(\ref{eq:A}) and (\ref{eq:id4}) were used in the third and fourth equalities, respectively.
As with the single-band quantum geometric tensor [Eq.~(\ref{eq:single-band-QGT})],
the second-order tensor decomposes into real and imaginary parts:
\begin{align}
Q_{mn}^{\mu\nu}
=
g_{mn}^{\mu\nu}
-\frac{i}{2} \Omega_{mn}^{\mu\nu},
\label{eq:SM-two-state-QGT}
\end{align}
where the real part
$
g_{mn}^{\mu\nu}
\equiv
{\rm Re} [ Q_{mn}^{\mu\nu} ]
$
is the two-state quantum metric, and
the imaginary part
$
\Omega_{mn}^{\mu\nu}
\equiv
-2{\rm Im} [ Q_{mn}^{\mu\nu} ]
$
is the two-state Berry curvature~\cite{Mitscherling2025}.

\subsubsection{Two-state quantum metric}

We rewrite $Q_{mn}^{\mu\nu*}$ in terms of the projectors:
\begin{align}
Q_{mn}^{\mu\nu*}
&=
-
{\rm Tr}
\Big[
(\partial_\nu P_n^\dagger)(\partial_\mu P_m^\dagger) P_n^\dagger
\Big]
=
-
{\rm Tr}
\Big[
(\partial_\nu P_n)(\partial_\mu P_m) P_n
\Big]
\\
&=
-
{\rm Tr}
\Big[
(\partial_\nu P_n)(- P_m \partial_\mu P_n)
\Big]
=
-
{\rm Tr}
\Big[
P_n (\partial_\nu P_m) (\partial_\mu P_n)
\Big],
\end{align}
where Eq.~(\ref{eq:id1}) was used in the last two equalities.
Using this expression, we obtain the projector representation of the two-state quantum metric:
\begin{align}
g_{mn}^{\mu\nu}
&=
{\rm Re} [ Q_{mn}^{\mu\nu} ]
=
\frac{Q_{mn}^{\mu\nu} + Q_{mn}^{\mu\nu*}}{2}
=
-\frac{1}{2}\mathrm{Tr}
\Big[
P_n (\partial_\mu P_m) (\partial_\nu P_n)
+
(\mu \leftrightarrow \nu)
\Big] .
\end{align}
This expression shows that the quantum metric is symmetric:
\begin{align}
g_{mn}^{\mu\nu} = g_{mn}^{\nu\mu}
~~~~{\rm and}~~~~
g_{mn}^{\mu\nu} = g_{nm}^{\mu\nu}.
\end{align}

\subsubsection{Two-state Berry curvature}

The projector representation of the two-state Berry curvature follows directly:
\begin{align}
\Omega_{mn}^{\mu\nu}
&=
-2{\rm Im} [ Q_{mn}^{\mu\nu} ]
=
i \left( Q_{mn}^{\mu\nu} - Q_{mn}^{\mu\nu*} \right)
=
-i
{\rm Tr}
\Big[
P_n(\partial_\mu P_m)(\partial_\nu P_n)
-(\mu\leftrightarrow\nu)
\Big].
\end{align}
This expression shows that the Berry curvature is antisymmetric:
\begin{align}
\Omega_{mn}^{\mu\nu} = - \Omega_{mn}^{\nu\mu}
~~~~{\rm and}~~~~
\Omega_{mn}^{\mu\nu} = - \Omega_{nm}^{\mu\nu}.
\end{align}

\subsubsection{Relation to the single-band geometric tensor}

The two-state tensors [Eq.~(\ref{eq:SM-two-state-QGT})] are closely related to the single-band tensors [Eq.~(\ref{eq:single-band-QGT})].
Specifically, the single-band tensors are obtained by summing the two-state tensors over a band index:
\begin{align}
Q_{n}^{\mu\nu}
&=
\sum_{m(\ne n)}
Q_{mn}^{\mu\nu}
=
-
\sum_{m(\ne n)}
{\rm Tr}
\Big[
P_n (\partial_\mu P_m) (\partial_\nu P_n)
\Big],
\\
g_{n}^{\mu\nu}
&=
\sum_{m(\ne n)}
g_{mn}^{\mu\nu}
=
-\frac{1}{2}
\sum_{m(\ne n)}
\mathrm{Tr}
\Big[
P_n (\partial_\mu P_m) (\partial_\nu P_n)
+
(\mu \leftrightarrow \nu)
\Big],
\\
\Omega_{n}^{\mu\nu}
&=
\sum_{m(\ne n)}
\Omega_{mn}^{\mu\nu}
=
-i
\sum_{m(\ne n)}
{\rm Tr}
\Big[
P_n(\partial_\mu P_m)(\partial_\nu P_n)
-(\mu\leftrightarrow\nu)
\Big],
\end{align}
which coincide with Eqs.~(\ref{eq:single-band-Q}), (\ref{eq:single-band-g}), and (\ref{eq:single-band-Omega}), respectively.

\subsection{Third- and fourth-order quantum geometric tensors}

Applying the same procedure as for the second-order tensor yields the following projector representations of higher-order geometric tensors:
\begin{align}
Q_{rmn}^{\rho\mu\nu}
=
A_{nr}^{\rho} A_{rm}^{\mu} A_{mn}^\nu
=
{\rm Tr}
\Big[
\hat{e}_{nr}^{\rho} \hat{e}_{rm}^\mu \hat{e}_{mn}^\nu
\Big]
&=
-i
{\rm Tr}
\Big[
P_n
(\partial_\rho P_r)
(\partial_\mu P_m)
(\partial_\nu P_n)
\Big],
\end{align}
\begin{align}
Q_{srmn}^{\sigma\rho\mu\nu}
=
A_{ns}^{\sigma} A_{sr}^{\rho} A_{rm}^{\mu} A_{mn}^\nu
=
{\rm Tr}
\Big[
\hat{e}_{ns}^{\sigma} \hat{e}_{sr}^{\rho} \hat{e}_{rm}^\mu \hat{e}_{mn}^\nu
\Big]
&=
\mathrm{Tr}
\Big[
P_n(\partial_\sigma P_s)(\partial_\rho P_r)(\partial_\mu P_m)(\partial_\nu P_n)
\Big].
\end{align}

\subsection{Geometric tensors with covariant derivatives}

Using the representations in Eqs.~(\ref{eq:A}) and (\ref{eq:Del-A}),
we readily obtain the projector representations of the remaining geometric tensors involving covariant derivatives.
\begin{align}
C_{mn}^{\mu(\rho;\nu)}
&=
A^\mu_{nm}\nabla_\rho A^\nu_{mn}
=
{\rm Tr}
\Big[
\hat{e}_{nm}^{\mu} \hat{\nabla}_\rho \hat{e}_{mn}^\nu
\Big]
\\
&=
-
{\rm Tr}
\Big\{
P_n (\partial_\mu P_m) P_m \Big[
(\partial_\rho P_m) (\partial_\nu P_n)
+
\partial_\rho \partial_\nu P_n
\Big]
\Big\}
\\
&=
-
{\rm Tr}
\Big\{
P_n (\partial_\mu P_m) \Big[
(\partial_\rho P_m) (\partial_\nu P_n)
+
\partial_\rho \partial_\nu P_n
\Big]
\Big\}
\end{align}
\begin{align}
B_{rmn}^{\mu\nu(\rho;\sigma)}
&=
A^\mu_{nr}A^\nu_{rm}\nabla_\rho A^\sigma_{mn}
=
{\rm Tr}
\Big[
\hat{e}_{nr}^{\mu} \hat{e}_{rm}^\nu \hat{\nabla}_\rho \hat{e}_{mn}^\sigma
\Big]
\\
&=
-i
{\rm Tr}
\Big\{
P_n (\partial_\mu P_r)
P_r (\partial_\nu P_m)
P_m
\Big[
(\partial_\rho P_m) (\partial_\sigma P_n)
+
\partial_\rho \partial_\sigma P_n
\Big]
\Big\}
\\
&=
-i
{\rm Tr}
\Big\{
P_n (\partial_\mu P_r)
P_r (\partial_\nu P_m)
\Big[
(\partial_\rho P_m) (\partial_\sigma P_n)
+
\partial_\rho \partial_\sigma P_n
\Big]
\Big\}
\\
&=
-i
{\rm Tr}
\Big\{
P_n (\partial_\mu P_r)
(\partial_\nu P_m)
\Big[
(\partial_\rho P_m) (\partial_\sigma P_n)
+
\partial_\rho \partial_\sigma P_n
\Big]
\Big\}
\end{align}
\begin{align}
D_{mn}^{(\rho;\mu)(\sigma;\nu)}
&=
(\nabla_\rho A^\mu_{nm})(\nabla_\sigma A^\nu_{mn})
=
{\rm Tr}
\Big[
(\hat{\nabla}_\rho \hat{e}_{nm}^\mu) (\hat{\nabla}_\sigma \hat{e}_{mn}^\nu)
\Big]
\\
&=
-
{\rm Tr}
\Big\{
P_n
\Big[
(\partial_\rho P_n) (\partial_\mu P_m)
+
\partial_\rho \partial_\mu P_m
\Big]
P_m
\Big[
(\partial_\sigma P_m) (\partial_\nu P_n)
+
\partial_\sigma \partial_\nu P_n
\Big]
\Big\}
\end{align}
In these derivations, we repeatedly used the identity in Eq.~(\ref{eq:id4}).

\section{Raman circular dichroism of two-magnon excitations}

\subsection{Raman operator}

In the two-magnon sector, the Raman operator is determined by second derivatives of the Hamiltonian matrix,
\begin{equation}
L_\mathbf{k}^{\mu\nu}
=
T_{\mathbf{k}}^\dagger
\bigl(\partial_{\mu}\partial_{\nu}H_{\mathbf{k}}\bigr)
T_{\mathbf{k}} .
\end{equation}
Although the $L$ matrix contains interband connections ($A$) and their derivatives ($\partial A$), these two objects transform differently under gauge transformations.
The covariant derivative $\nabla_\mu A_{mn}^\nu({\bf k}) =
\left( \partial_\mu -i A_{mm}^\mu + i A_{nn}^\mu \right)
A^\nu_{mn}$, however, transforms in exactly the same way as the interband connection $A_{mn}^\nu({\bf k})$, i.e.,
\begin{equation}
\left[
\begin{array}{c}
A_{mn}^\nu({\bf k})
\\
\nabla_\mu A_{mn}^\nu({\bf k})
\end{array}
\right]
\rightarrow
\left[
\begin{array}{c}
A_{mn}^\nu({\bf k})
\\
\nabla_\mu A_{mn}^\nu({\bf k})
\end{array}
\right]
e^{i[-\phi_m({\bf k})+\phi_n({\bf k})]}
\end{equation}
under the gauge transformation $T_{\bf k} \rightarrow T_{\bf k}~{\rm diag}\Big[ e^{i\phi_1({\bf k})}, \cdots , e^{i\phi_{2N}({\bf k})}\Big]$.
It is therefore natural to express the $L$ matrix in terms of $A$ and $\nabla A$.

We first expand the $L$ matrix in terms of $A$, $AA$, and $\partial A$.
Differentiating Eq.~(\ref{eq:dH}) and symmetrizing over the indices $\mu$ and $\nu$ gives
\begin{align}
L_{\bf k}^{\mu\nu}
&=
\partial_{\mu} \partial_{\nu} D_{\bf k}
\nonumber\\
&+
\Big[ i(\partial_{\mu} D_{\bf k}) A_{\bf k}^\nu + i(\partial_{\nu} D_{\bf k}) A_{\bf k}^\mu\Big]+{\rm H.c.}
\nonumber\\
&+
\Big[ -\frac{1}{2}D_{\bf k} (A_{\bf k}^\mu A_{\bf k}^\nu + A_{\bf k}^\nu A_{\bf k}^\mu)
+ A_{\bf k}^{\mu\dagger} D_{\bf k} A_{\bf k}^{\nu}
\Big]+{\rm H.c.}
\nonumber\\
&+
\Big[
\frac{i}{2} D_{\bf k} (\partial_{\mu} A_{\bf k}^\nu + \partial_{\nu} A_{\bf k}^\mu)
\Big]+{\rm H.c.}
\end{align}
We focus on the off-diagonal elements $L^{\mu\nu}_{mn}$ ($m \neq n$) and rewrite $\partial A$ in terms of $\nabla A$ and $AA$ using
\begin{align}
\partial_\mu A^\nu_{mn}
=
\nabla_\mu A_{mn}^\nu
+
i A_{mm}^\mu A^\nu_{mn}
-
i A_{nn}^\mu A^\nu_{mn} .
\end{align}
This gives
\begin{align}
L_{mn}^{\mu\nu}
=\Big( i v^\mu_{mn}A^\nu_{mn}
+ i d_{mn} \nabla_\mu A^\nu_{mn}
+ \sum_{r(\ne m,n)} d_{mnr}A^\mu_{mr}A^\nu_{rn} \Big)
+ \Big( \mu \leftrightarrow \nu \Big),
\label{eq:L-mat}
\end{align}
where
\begin{align}
v^\mu_{mn} &=\partial_\mu\Big(d_m - [I_B]_m [I_B]_n d_n\Big),
\\
d_{mn} &=\frac{1}{2}\Big(d_m - [I_B]_m [I_B]_n d_n\Big),
\\
d_{mnr} &=\frac{1}{2}\Big(- d_m - [I_B]_m [I_B]_n d_n\Big) + [I_B]_m [I_B]_r d_r,
\end{align}
with $d_r=[D_{\bf k}]_r$.
The matrix element $L^{\mu\nu}_{mn}$ is thus organized into three components, $A$, $\nabla A$, and $AA$, all of which obey the same gauge transformation rule.

\subsection{Two-magnon Raman circular dichroism}

Within our framework, the two-magnon RCD intensity is
\begin{align}
I^{\rm 2M}_{\rm RCD}(\omega)
&=
I_{LR}^{\rm 2M}(\omega)
-
I_{RL}^{\rm 2M}(\omega)
=\frac{2\pi}{V}
\sum_{{\bf k}}\sum_{m=1}^{N}\sum_{n=N+1}^{2N}
{\rm Im} \Big[ (L^{xx}_{nm}-L^{yy}_{nm}) L^{xy}_{mn} \Big]
\delta(\omega-\omega_{m{\bf k}}-\omega_{n-N,-{\bf k}}) .
\end{align}
Substituting Eq.~(\ref{eq:L-mat}) into this expression and grouping the terms according to the six types of geometric tensors yields six distinct contributions to the RCD intensity:
\begin{align}
I^{\rm 2M}_{\rm RCD}(\omega)
=
I^{\rm 2M}_{\rm RCD}\{AA\}+I^{\rm 2M}_{\rm RCD}\{AAA\}+I^{\rm 2M}_{\rm RCD}\{AAAA\}
+I^{\rm 2M}_{\rm RCD}\{A\nabla A\}+I^{\rm 2M}_{\rm RCD}\{AA\nabla A\}+I^{\rm 2M}_{\rm RCD}\{\nabla A\nabla A\} .
\end{align}
The matrix elements associated with these six contributions are given explicitly below in terms of the geometric tensors.

\begin{align}
{\rm Im} \Big[
(L^{xx}_{nm}-L^{yy}_{nm}) L^{xy}_{mn}
\Big]_{AA}
=
\Big( v^x_{nm} v^x_{mn} +v^y_{nm} v^y_{mn} \Big) \Omega^{xy}_{mn}
\end{align}

\begin{align}
{\rm Im}\Big[
(L^{xx}_{nm}-L^{yy}_{nm}) L^{xy}_{mn}
\Big]_{AAA}
&=
\sum_{r(\ne m,n)}
2 d_{mnr}
\Big[
v^x_{nm}{\rm Re}( Q_{mrn}^{xxy} + Q_{mrn}^{xyx} )
-
v^y_{nm}{\rm Re}( Q_{mrn}^{yxy} + Q_{mrn}^{yyx} )
\Big]
\\
&+
\sum_{r(\ne m,n)}
2 d_{nmr}
\Big[
v^x_{mn}{\rm Re}( Q_{rmn}^{xxy} - Q_{rmn}^{yyy} )
+
v^y_{mn}{\rm Re}( Q_{rmn}^{xxx} - Q_{rmn}^{yyx} )
\Big]
\end{align}

\begin{align}
{\rm Im} \Big[
(L^{xx}_{nm}-L^{yy}_{nm}) L^{xy}_{mn}
\Big]_{AAAA}
&=
\sum_{r\ne s(\ne m,n)}
2 d_{nmr}d_{mns} {\rm Im} \Big[ Q_{rmsn}^{xxxy} + Q_{rmsn}^{xxyx} - Q_{rmsn}^{yyxy} - Q_{rmsn}^{yyyx} \Big]
\\
&+
\sum_{r(\ne m,n)}
d_{nmr}d_{mnr}
\Big[
-
( g_{mr}^{xx} +  g_{mr}^{yy} ) \Omega_{rn}^{xy}
-
\Omega_{mr}^{xy} ( g_{rn}^{xx} + g_{rn}^{yy} )
\Big]
\end{align}

\begin{align}
{\rm Im} \Big[
(L^{xx}_{nm}-L^{yy}_{nm}) L^{xy}_{mn}
\Big]_{A\nabla A}
&=
2
v^x_{nm} d_{mn}
{\rm Im}\Big[
-C^{x(x;y)}_{mn}-C^{x(y;x)}_{mn}
\Big]
+2
d_{nm} v^x_{mn}
{\rm Im}\Big[
-C^{y(x;x)}_{nm}+C^{y(y;y)}_{nm}
\Big]
\\
&
+2
v^y_{nm} d_{mn}
{\rm Im}\Big[
C^{y(x;y)}_{mn}+C^{y(y;x)}_{mn}
\Big]
+2
d_{nm} v^y_{mn}
{\rm Im}\Big[
-C^{x(x;x)}_{nm}+C^{x(y;y)}_{nm}
\Big]
\end{align}

\begin{align}
{\rm Im} \Big[
(L^{xx}_{nm}-L^{yy}_{nm}) L^{xy}_{mn}
\Big]_{AA\nabla A}
&=
\sum_{r(\ne m,n)}
2 d_{nmr} d_{mn}
{\rm Re}\Big[
B^{xx(x;y)}_{rmn}+B^{xx(y;x)}_{rmn}-B^{yy(x;y)}_{rmn}-B^{yy(y;x)}_{rmn}
\Big]
\\
&+
\sum_{r(\ne m,n)}
2 d_{nm} d_{mnr}
{\rm Re}\Big[
B^{xy(x;x)}_{rnm}+B^{yx(x;x)}_{rnm}-B^{xy(y;y)}_{rnm}-B^{yx(y;y)}_{rnm}
\Big]
\end{align}

\begin{align}
{\rm Im} \Big[
(L^{xx}_{nm}-L^{yy}_{nm}) L^{xy}_{mn}
\Big]_{\nabla A\nabla A}
&=
2 d_{nm} d_{mn}
{\rm Im}\Big[
-
D^{(x;x)(x;y)}_{mn}
-
D^{(x;x)(y;x)}_{mn}
+
D^{(y;y)(x;y)}_{mn}
+
D^{(y;y)(y;x)}_{mn}
\Big]
\end{align}

\end{document}